\documentclass[12pt]{iopart}
\usepackage{graphicx}

\begin{document}

\title[Entanglement in open quantum systems]{Entanglement in two-mode continuous variable\\
open quantum systems}

\author{Aurelian Isar}

\address{National Institute of Physics and Nuclear Engineering,
P.O.Box MG-6, Bucharest-Magurele, Romania}
\ead{isar@theory.nipne.ro}
\begin{abstract}

In the framework of the theory of open systems based on completely
positive quantum dynamical semigroups,
we give a description of the continuous-variable entanglement for a system consisting of
two noninteracting modes embedded in a thermal environment.
By using the Peres-Simon necessary and sufficient criterion for separability of two-mode
Gaussian states, we describe the evolution of entanglement in terms
of the covariance matrix for Gaussian input
states. For all values of the temperature of the thermal reservoir, an initial separable Gaussian state remains separable for all times. In the case of an entangled initial Gaussian state, entanglement suppression (entanglement sudden death) takes place, for non-zero temperatures of the environment. Only for a zero temperature of the thermal bath the initial entangled state remains entangled for finite times.
We also show that, independent of its type -- separable or entangled, the
initial state evolves asymptotically to an
equilibrium state which is always separable.
\end{abstract}


\section{Introduction}

In recent years there is an increasing interest in using continuous variable (CV) systems in applications of quantum information processing, communication and computation \cite{bra1}. The realization of quantum information processing tasks depends on the generation and manipulation of nonclassical entangled states of CV systems. A full characterization of the nonclassical properties of entangled states of CV systems exists, at present, only for the class of Gaussian states. In this special case there exist necessary and sufficient criteria of entanglement \cite{sim,dua} and quantitative entanglement measures \cite{vid,gie}. In quantum information theory of CV systems, Gaussian states, in particular two-mode Gaussian states, play a key role since they can be easily created and controlled experimentally.

Implementation of quantum communication and computation encounters the difficulty that any realistic quantum system cannot be isolated and it always has to interact with its environment. Quantum coherence and entanglement of quantum systems are inevitably influenced during their interaction with the external environment. As a result of the irreversible and uncontrollable phenomenon of quantum decoherence, the purity and entanglement of
quantum states are in most cases degraded. Practically, compared with the discrete variable entangled states, the CV entangled states may be more efficient because they are less affected by decoherence.

Due to the unavoidable interaction with the environment, any pure quantum state evolves into a mixed state and to describe realistically CV quantum information processes it is necessary to take decoherence and dissipation into consideration. Decoherence and dynamics of quantum entanglement in CV open systems have been intensively studied in the last years \cite{dua1,oli,pra,dod1,dod2,avd,ser2,ser1,ben1,ban,mch,man,jan,aphysa,aeur,arus1}. When two systems are immersed in an environment, then, in addition to and at the same time with the quantum decoherence phenomenon, the environment can also generate a quantum entanglement of the two systems and therefore an additional mechanism to correlate them \cite{ben1,vvd1,ben2,paz1,paz2}.

In this paper we study, in the framework of the theory of open systems based on completely positive quantum dynamical semigroups, the dynamics of the CV entanglement of two uncoupled modes (two uncoupled harmonic oscillators) interacting with a common thermal environment. The initial state of the subsystem is taken of Gaussian form and the evolution under the quantum dynamical semigroup assures the preservation in time of the Gaussian form of the state. We have studied previously \cite{ascri,aosid} the evolution of the entanglement of two identical harmonic oscillators interacting with a general environment, characterized by general diffusion and dissipation coefficients. We obtained that, depending on the values of these coefficients, the state keeps for all times its initial type:
separable or entangled. In other cases, entanglement generation, entanglement sudden death or a periodic collapse and revival of entanglement take place.

The paper is organized as follows. In Sec. 2 we write the Markovian master equation in the Heisenberg representation for two uncoupled harmonic oscillators interacting with a general environment and the evolution equation for the covariance matrix. For this equation we give its general solution, i.e. we derive the variances and covariances of coordinates and momenta corresponding to a generic two-mode Gaussian state. By using the Peres-Simon necessary and sufficient condition for separability of two-mode Gaussian states \cite{sim,per}, we investigate in Sec. 3 the dynamics of entanglement for the considered subsystem. We show that for all values of the temperature of the thermal reservoir, an initial separable Gaussian state remains separable for all times. In the case of an entangled initial Gaussian state, entanglement suppression (entanglement sudden death) takes place, for non-zero temperatures of the environment. Only for a zero temperature of the thermal bath the initial entangled state remains entangled for all finite times, but in the limit of infinite time it evolves asymptotically to an equilibrium state which is always separable.  We analyze also the time evolution of the logarithmic negativity, which characterizes the degree of entanglement of the quantum state. A summary is given in Sec. 4.

\section{Equations of motion for two harmonic oscillators}

We study the dynamics
of the subsystem composed of two non-interacting oscillators in weak interaction with a thermal environment. In the axiomatic formalism
based on completely positive quantum dynamical semigroups, the irreversible time
evolution of an open system is described by the following general quantum Markovian master equation
for an operator $A$ in the Heisenberg representation ($\dagger$ denotes Hermitian conjugation) \cite{lin,rev}:
\begin{eqnarray}{dA\over dt}={\rmi\over \hbar}[H,A]+{1\over
2\hbar}\sum_j(V_j^{\dagger}[A,
V_j]+[V_j^{\dagger},A]V_j).\label{masteq}\end{eqnarray}
Here, $H$ denotes the Hamiltonian of the open system
and the operators $V_j, V_j^\dagger,$ defined on the Hilbert space of $H,$
represent the interaction of the open system
with the environment.

We are interested in the set of Gaussian states, therefore we introduce such quantum
dynamical semigroups that preserve this set during time evolution of the system and in this case our model represents a Gaussian noise channel.
Consequently $H$ is
taken to be a polynomial of second degree in the coordinates
$x,y$ and momenta $p_x,p_y$ of the two quantum oscillators and
$V_j,V_j^{\dagger}$ are taken polynomials of first degree
in these canonical observables. Then in the linear space
spanned by the coordinates and momenta there exist only four
linearly independent operators $V_{j=1,2,3,4}$ \cite{san}: \begin{eqnarray}
V_j=a_{xj}p_x+a_{yj}p_y+b_{xj}x+b_{yj}y,\end{eqnarray} where
$a_{xj},a_{yj},b_{xj},b_{yj}$ are complex coefficients.
The Hamiltonian of the two uncoupled non-resonant harmonic
oscillators of identical mass $m$ and frequencies $\omega_1$ and $\omega_2$ is
\begin{eqnarray}
H={1\over 2m}(p_x^2+p_y^2)+\frac{m}{2}(\omega_1^2 x^2+\omega_2^2 y^2).\end{eqnarray}

The fact that the evolution is given by a dynamical semigroup
implies the positivity of the matrix formed by the
scalar products of the four vectors $ {\bf a}_x, {\bf b}_x,
{\bf a}_y, {\bf b}_y,$ whose entries are the components $a_{xj},b_{xj},a_{yj},b_{yj},$
respectively.
We take this matrix of the following form, where
all coefficients $D_{xx}, D_{xp_x},$... and $\lambda$ are real quantities (we
put from now on $\hbar=1$):
\begin{eqnarray} \left(\matrix{D_{xx}&- D_{xp_x} - i \frac{\lambda}{2}& D_{xy}&-
D_{xp_y} \cr - D_{xp_x} + i \frac{\lambda}{2}&D_{p_x p_x}&-
D_{yp_x}&D_{p_x p_y} \cr D_{xy}&- D_{y p_x}&D_{yy}&- D_{y p_y}
- i \frac{\lambda}{2} \cr - D_{xp_y} &D_{p_x p_y}&- D_{yp_y} + i
\frac{\lambda}{2}&D_{p_y p_y}}\right).\label{coef} \end{eqnarray} It follows that
the principal minors of this matrix are positive or zero. From
the Cauchy-Schwarz inequality the following relations hold for the
coefficients defined in Eq. (\ref{coef}): \begin{eqnarray}
D_{xx}D_{p_xp_x}-D^2_{xp_x}\ge\frac{\lambda^2}{4},~
D_{yy}D_{p_yp_y}-D^2_{yp_y}\ge\frac{\lambda^2}{4},\nonumber\\
D_{xx}D_{yy}-D^2_{xy}\ge0,~
D_{p_xp_x}D_{p_yp_y}-D^2_{p_xp_y}\ge 0, \nonumber \\
D_{xx}D_{p_yp_y}-D^2_{xp_y}\ge 0,~D_{yy}D_{p_xp_x}-D^2_{yp_x}\ge 0.
\label{coefineq}\end{eqnarray}

We introduce the following $4\times 4$ bimodal covariance matrix:
\begin{eqnarray}\sigma(t)=\left(\matrix{\sigma_{xx}(t)&\sigma_{xp_x}(t) &\sigma_{xy}(t)&
\sigma_{xp_y}(t)\cr \sigma_{xp_x}(t)&\sigma_{p_xp_x}(t)&\sigma_{yp_x}(t)
&\sigma_{p_xp_y}(t)\cr \sigma_{xy}(t)&\sigma_{yp_x}(t)&\sigma_{yy}(t)
&\sigma_{yp_y}(t)\cr \sigma_{xp_y}(t)&\sigma_{p_xp_y}(t)&\sigma_{yp_y}(t)
&\sigma_{p_yp_y}(t)}\right).\label{covar} \end{eqnarray}
We can transform the problem
of solving the master equation for the operators in Heisenberg representation into a problem of solving first-order in time,
coupled linear differential equations for the covariance matrix elements.
Namely, from Eq. (\ref{masteq}) we obtain the following system of equations for the quantum correlations of the canonical observables ($\rm T$ denotes the transposed matrix) \cite{san}:
\begin{eqnarray}{d \sigma(t)\over
dt} = Y \sigma(t) + \sigma(t) Y^{\rm T}+2 D,\label{vareq}\end{eqnarray} where
\begin{eqnarray} Y=\left(\matrix{ -\lambda&1/m&0 &0\cr -m\omega_1^2&-\lambda&0&
0\cr 0&0&-\lambda&1/m \cr 0&0&-m\omega_2^2&-\lambda}\right),\\
D=\left(\matrix{
D_{xx}& D_{xp_x} &D_{xy}& D_{xp_y} \cr D_{xp_x}&D_{p_x p_x}&
D_{yp_x}&D_{p_x p_y} \cr D_{xy}& D_{y p_x}&D_{yy}& D_{y p_y}
\cr D_{xp_y} &D_{p_x p_y}& D_{yp_y} &D_{p_y p_y}} \right).\end{eqnarray}
The time-dependent
solution of Eq. (\ref{vareq}) is given by \cite{san}
\begin{eqnarray}\sigma(t)= M(t)[\sigma(0)-\sigma(\infty)] M^{\rm
T}(t)+\sigma(\infty),\label{covart}\end{eqnarray} where the matrix $M(t)=\exp(Yt)$ has to fulfill
the condition $\lim_{t\to\infty} M(t) = 0.$
In order that this limit exists, $Y$ must only have eigenvalues
with negative real parts. The values at infinity are obtained
from the equation \begin{eqnarray}
Y\sigma(\infty)+\sigma(\infty) Y^{\rm T}=-2 D.\label{covarinf}\end{eqnarray}

\section{Dynamics of two-mode continuous variable entanglement}

A well-known sufficient condition for inseparability is the
so-called Peres-Horodecki criterion \cite{per,hor} which is based on
the observation that the non-completely positive nature of the
partial transposition operation of the density matrix for a bipartite system (transposition with respect to degrees of freedom of one subsystem only) may turn an inseparable state
into a nonphysical state. The signature of this non-physicality,
and thus of quantum entanglement, is the appearance of a negative
eigenvalue in the eigenspectrum of the partially transposed
density matrix of a bipartite system. The characterization of the separability
of CV states using second-order moments of quadrature operators was given in Refs. \cite{sim,dua}. For Gaussian states,
whose statistical properties are fully characterized by just
second-order moments, this criterion was proven to be necessary
and sufficient: A Gaussian CV state is separable if and only if the partial transpose
of its density matrix is non-negative [positive partial
transpose (PPT) criterion].

The two-mode Gaussian state is entirely specified by its
covariance matrix (\ref{covar}), which is a real,
symmetric and positive matrix with the following block
structure:
\begin{eqnarray}
\sigma(t)=\left(\begin{array}{cc}A&C\\
C^{\rm T}&B \end{array}\right),\label{cm}
\end{eqnarray}
where $A$, $B$ and $C$ are $2\times 2$ Hermitian matrices. $A$
and $B$ denote the symmetric covariance matrices for the
individual reduced one-mode states, while the matrix $C$
contains the cross-correlations between modes. When these correlations
have non-zero values, then the states with $\det C\ge 0$ are
separable states, but for $\det C <0$ it may be possible that
the states are entangled.

The $4\times 4$ covariance matrix (\ref{cm}) (where all first moments
have been set to zero by means of local unitary operations which do not affect the entanglement) contains
four local symplectic invariants in form of the determinants
of the block matrices $A, B, C$ and covariance matrix $\sigma.$ Based on the above invariants
Simon \cite{sim} derived a PPT criterion for bipartite Gaussian
CV states: the necessary and sufficient criterion for separability is
$S(t)\ge 0,$ where \begin{eqnarray} S(t)\equiv\det A \det B+(\frac{1}{4} -|\det
C|)^2\nonumber\\
- {\rm Tr}[AJCJBJC^{\rm T}J]- \frac{1}{4}(\det A+\det B)
\label{sim1}\end{eqnarray} and $J$ is the $2\times 2$ symplectic matrix
\begin{eqnarray}
J=\left(\begin{array}{cc}0&1\\
-1&0\end{array}\right).
\end{eqnarray}

\subsection{Time evolution of entanglement and logarithmic negativity}

We suppose that the asymptotic state of the considered open system is a Gibbs state corresponding to two independent quantum harmonic oscillators in thermal equilibrium at temperature $T.$ Then the quantum diffusion coefficients have the following form \cite{rev}:
\begin{eqnarray}m\omega_1 D_{xx}=\frac{D_{p_xp_x}}{m\omega_1}=\frac{\lambda}{2}\coth\frac{\omega_1}{2kT},\nonumber\\
m\omega_2 D_{yy}=\frac{D_{p_yp_y}}{m\omega_2}=\frac{\lambda}{2}\coth\frac{\omega_2}{2kT},\label{envcoe}\\
D_{xp_x}=D_{yp_y}=D_{xy}=D_{p_xp_y}=D_{xp_y}=D_{yp_x}=0.\nonumber\end{eqnarray}

The elements of the covariance matrix depend on $Y$ and $D$ and can be
calculated from Eqs. (\ref{covart}), (\ref{covarinf}). Solving for the time evolution of
the covariance matrix elements, we can obtain the entanglement
dynamics through the computation of the Simon criterion.

For Gaussian states, the measures of entanglement of bipartite systems are based on some invariants constructed from the elements of the covariance matrix \cite{oli,avd,ovm}. In order to quantify the degrees of entanglement of the infinite-dimensional bipartite system states of the two oscillators it is suitable to use the logarithmic negativity.  For a Gaussian density operator, the logarithmic negativity is completely defined by the symplectic spectrum of the partial transpose of the covariance matrix. It is given by
$
E_N=-\log_2 2\tilde\nu_-,
$
where $\tilde\nu_-$ is the smallest of the two symplectic eigenvalues of the partial transpose $\tilde{{\sigma}}$ of the 2-mode covariance matrix $\sigma:$
\begin{eqnarray}2\tilde{\nu}_{\mp}^2 = \tilde{\Delta}\mp\sqrt{\tilde{\Delta}^2
-4\det\sigma}
\end{eqnarray}
and $ \tilde\Delta$ is the symplectic invariant (seralian), given by
$ \tilde\Delta=\det A+\det B-2\det C.$

In our model, the logarithmic negativity is calculated as \begin{eqnarray}E_N(t)=-\frac{1}{2}\log_2[4f(\sigma(t))], \end{eqnarray} where \begin{eqnarray}f(\sigma(t))=\frac{1}{2}(\det A +\det
B)-\det C\nonumber\\
-\left({\left[\frac{1}{2}(\det A+\det B)-\det
C\right]^2-\det\sigma(t)}\right)^{1/2}.\end{eqnarray}
It determines the strength of entanglement for $E_N(t)>0,$ and if $E_N(t)\le 0,$ then the state is
separable.

In the following, we analyze the dependence of the Simon function $S(t)$  and of the logarithmic negativity $E_N(t)$ on time $t$ and temperature $T$ in the case of a thermal bath, when the diffusion coefficients are given by Eqs. (\ref{envcoe}). We consider two types of the initial Gaussian state: 1) separable and 2) entangled.

1) In Figure 1 we represent the dependence of the function $S(t)$ on time $t$ and temperature $T$ for a separable initial Gaussian state, with the two modes initially prepared in their single-mode squeezed states (unimodal squeezed state). Therefore the initial covariance matrix is taken of the form
\begin{eqnarray}\sigma(0)=\frac{1}{2}\left(\matrix{\cosh r&\sinh r &0&0\cr
\sinh r&\cosh r&0&0\cr
0&0&\cosh r&\sinh r\cr
0&0&\sinh r&\cosh r}\right),\label{ini1} \end{eqnarray}
where $r$ denotes the squeezing parameter. We notice that $S(t)$ becomes strictly positive after the initial moment of time ($S(0)=0),$ so that the initial separable state remains separable for all values of the temperature $T$ and for all times.

\begin{figure}
\centerline
{
\includegraphics{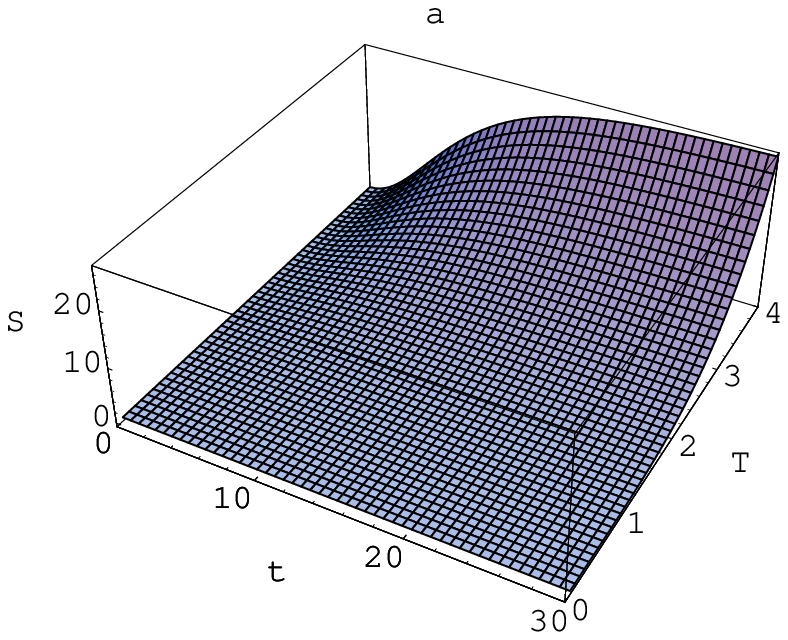}
\includegraphics{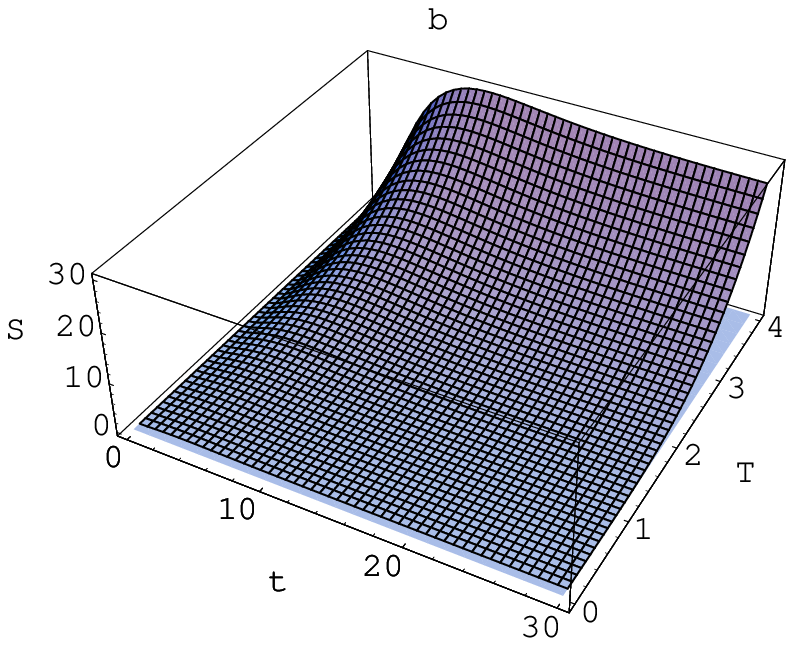}}
\caption{Separability function $S$ versus time $t$
and environment temperature $T$ for $\lambda=0.1, \omega_1=1, \omega_2=3$ and separable initial uni-modal squeezed state with squeezing parameter: a) $r=1/2;$ b) $r=2.$ We take $m=\hbar=k=1.$
\label{fig:1}}
\end{figure}

2) The evolution of an entangled initial state is illustrated in Figures 2 and 3, where we represent the dependence of the function $S(t)$ and logarithmic negativity $E_N(t)$ on time $t$ and temperature $T$ for an entangled initial Gaussian state, taken of the form of a two-mode vacuum squeezed state, with the initial covariance matrix given by
\begin{eqnarray}\sigma(0)=\frac{1}{2}\left(\matrix{\cosh r&0&\sinh r &0\cr
0&\cosh r&0&-\sinh r\cr
\sinh r&0&\cosh r&0\cr
0&-\sinh r&0&\cosh r}\right).\label{ini2} \end{eqnarray}
We observe that for a non-zero temperature $T,$ at certain finite moment of time, which depends on $T,$ $S(t)$ becomes zero and therefore the state becomes separable. This is the so-called phenomenon of entanglement sudden death. It corresponds to the finite moment of time when the logarithmic negativity becomes zero. This phenomenon is in contrast to the quantum decoherence, during which the loss of quantum coherence is usually gradual \cite{aphysa,arus}. For $T=0,$ $S(t)$ remains strictly negative for finite times and tends asymptotically to 0 for $t\to \infty.$ Therefore, only for zero temperature of the thermal bath the initial entangled state remains entangled for finite times and this state tends asymptotically to a separable one for infinitely large times. We notice also that the dissipation favorizes the phenomenon of entanglement sudden death -- with increasing the dissipation parameter $\lambda,$ the entanglement sudden death happens earlier.

\begin{figure}
\centerline
{
\includegraphics{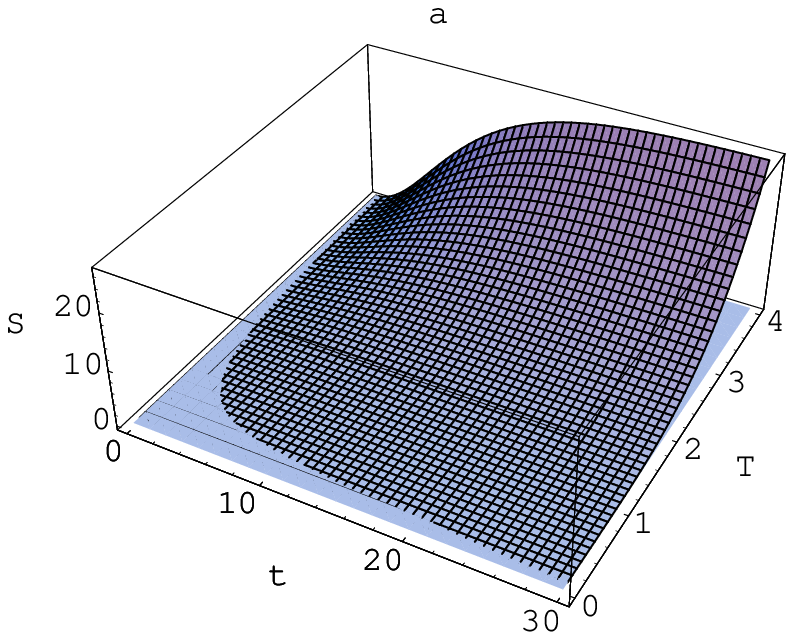}
\includegraphics{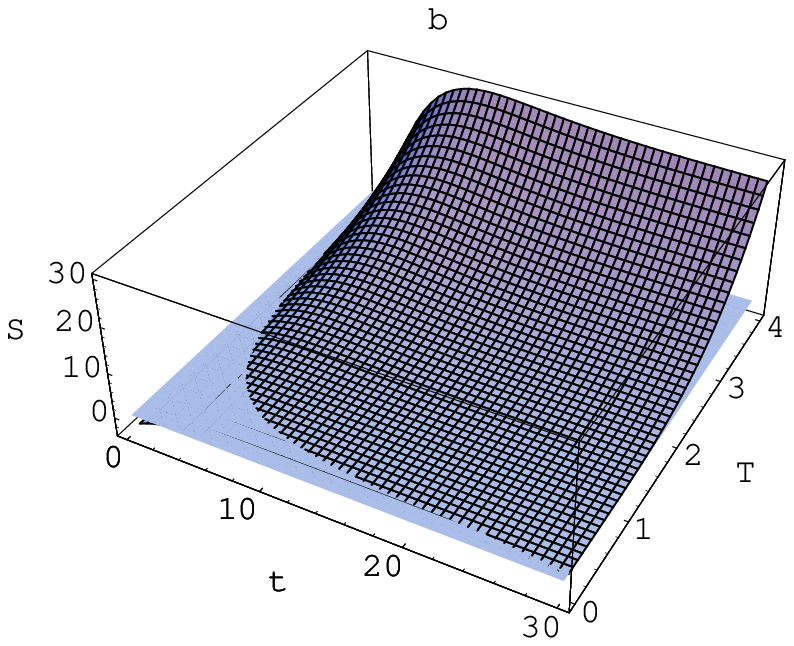}}
\caption{Separability function $S$ versus time $t$
and environment temperature $T$ for $\lambda=0.1, \omega_1=1, \omega_2=3$ and entangled initial vacuum squeezed state with squeezing parameter: a) $r=1/2;$ b) $r=2.$ We take $m=\hbar=k=1.$
\label{fig:2}}
\end{figure}

The dynamics of entanglement of the two oscillators depends strongly on the initial states and the coefficients describing the interaction of the system with the thermal environment (dissipation constant and temperature). As expected, the logarithmic negativity has a behaviour similar to that one of the Simon function in what concerns the characteristics of the state of being separable or entangled \cite{ascri,aosid,arus,aijqi}.

\begin{figure}
\centerline
{
\includegraphics{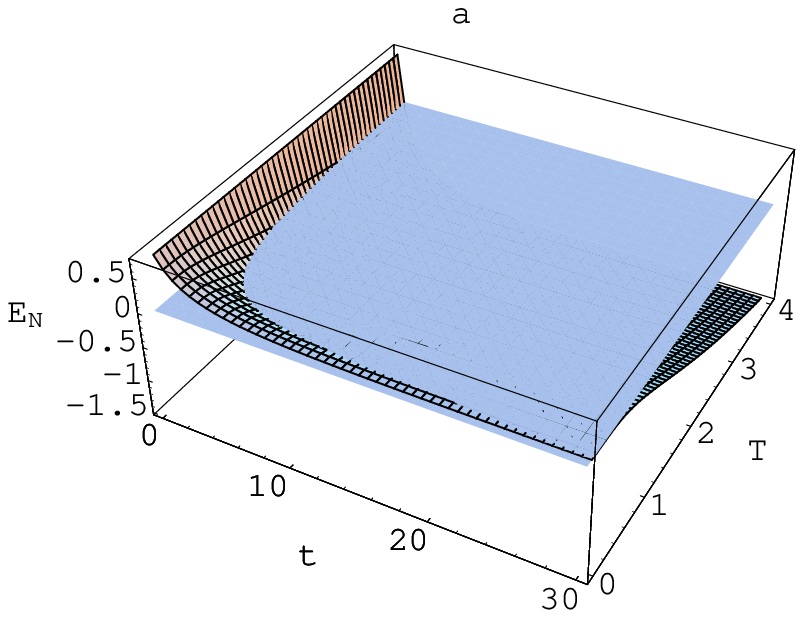}
\includegraphics{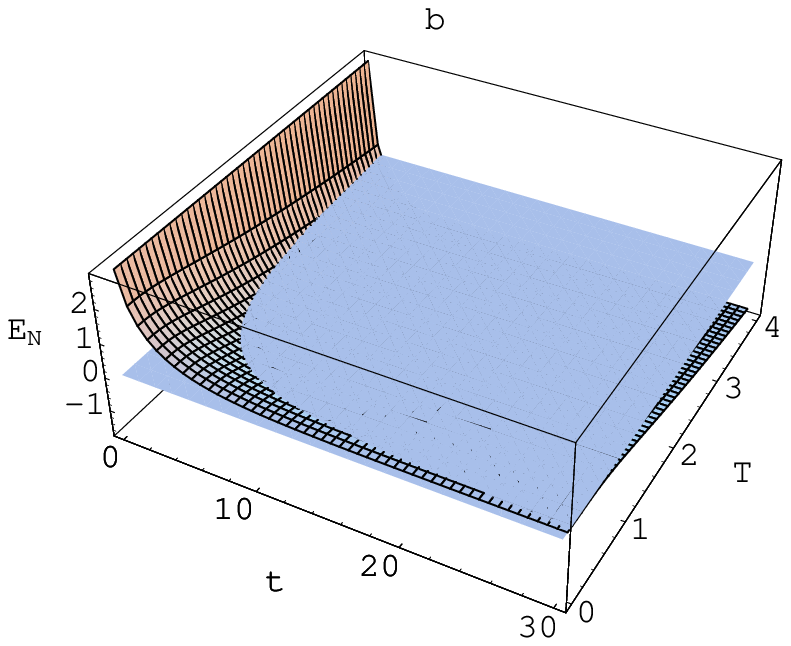}}
\centerline
{\includegraphics{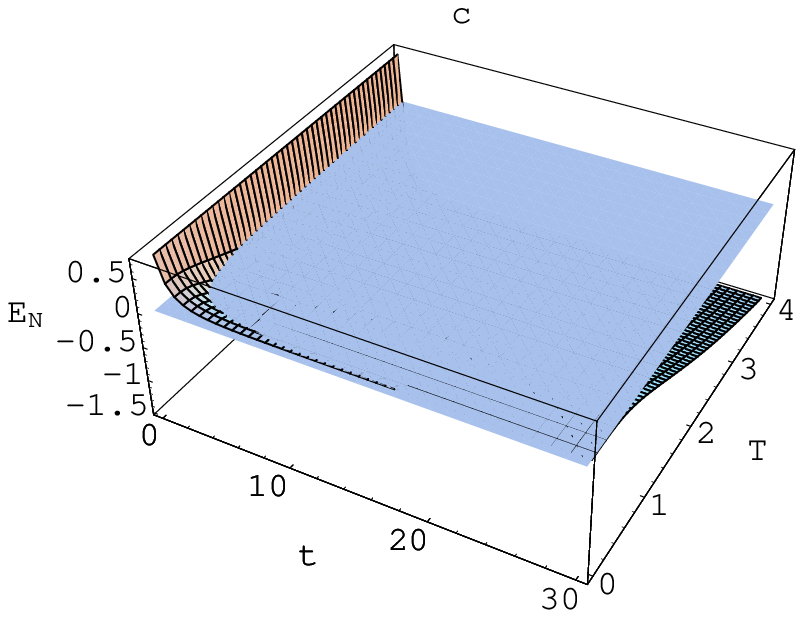}
\includegraphics{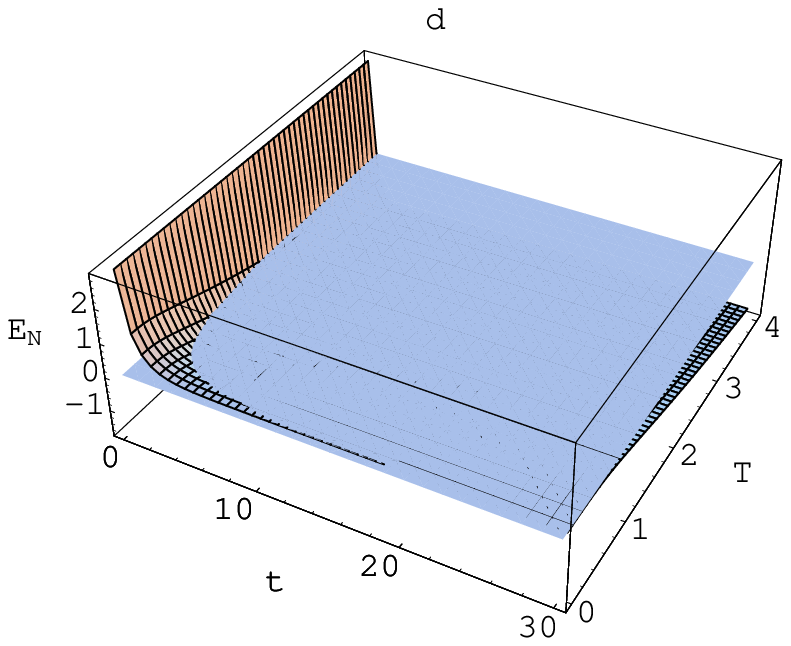}}
\caption{Logarithmic negativity $E_N$ versus time $t$
and environment temperature $T$ for $\omega_1=1, \omega_2=3$ and entangled initial vacuum squeezed state with squeezing parameter: a) $\lambda=0.1, r=1/2;$ b) $\lambda=0.1, r=2;$ c) $\lambda=0.3, r=1/2;$ d) $\lambda=0.3, r=2.$ We take $m=\hbar=k=1.$
\label{fig:3}}
\end{figure}

\subsection{Asymptotic entanglement}

On general grounds, one expects that the effects of
decoherence is dominant
in the long-time regime, so that no quantum correlations (entanglement) is expected to be left at infinity.
Indeed, using the diffusion coefficients given by Eqs. (\ref{envcoe}), we obtain from Eq. (\ref{covarinf}) the
following elements of the asymptotic matrices $A(\infty)$ and $B(\infty):$
\begin{eqnarray} m\omega_1\sigma_{xx}(\infty)=\frac{\sigma_{p_xp_x}(\infty)}{m\omega_1}=\frac{1}{2}\coth\frac{\omega_1}{2kT}, ~~~\sigma_{xp_x}(\infty)=0,\nonumber\\
m\omega_2\sigma_{yy}(\infty)=\frac{\sigma_{p_yp_y}(\infty)}{m\omega_2}=\frac{1}{2}\coth\frac{\omega_2}{2kT}, ~~~\sigma_{yp_y}(\infty)=0
\label{varinf} \end{eqnarray}
and of the entanglement matrix $C(\infty):$
\begin{eqnarray}\sigma_{xy} (\infty) =
\sigma_{xp_y}(\infty)=
\sigma_{yp_x}(\infty)=\sigma_{p_xp_y} (\infty) =0.\end{eqnarray}
Then the Simon expression (\ref{sim1}) takes the following form in the limit of large times,
\begin{eqnarray} S(\infty)=
\frac{1}{16}(\coth^2\frac{\omega_1}{2kT}-1)(\coth^2\frac{\omega_2}{2kT}-1),\label{sim2}\end{eqnarray}
and, correspondingly, the equilibrium asymptotic state is always separable in the case of two non-interacting harmonic oscillators immersed in a common thermal reservoir.

In Refs. \cite{ascri,aosid,arus,aijqi,arus2} we described the dependence of the logarithmic negativity $E_N(t)$ on time and mixed diffusion coefficient for two harmonic oscillators interacting with a general environment. In the present case of a thermal bath, the asymptotic logarithmic negativity is given by (for $\omega_1\le\omega_2$)
\begin{eqnarray} E_N(\infty)=-\log_2\coth\frac{\omega_2}{2kT}.\end{eqnarray}
It depends only on the temperature, and does not depend on the initial Gaussian state. $E_N(\infty)<0$ for $T\neq 0$ and $E_N(\infty)=0$ for $T=0,$ and this confirms the previous statement that the asymptotic state is always separable.

\section{Summary}

In the framework of the theory of open quantum systems based on completely positive quantum dynamical semigroups, we investigated the Markovian dynamics of the quantum entanglement for a subsystem
composed of two noninteracting modes embedded in a thermal bath. We have presented and discussed the influence of the environment on the entanglement dynamics for different initial states.
By using the Peres-Simon
necessary and sufficient condition for separability of two-mode
Gaussian states, we have described the evolution of entanglement in terms
of the covariance matrix for Gaussian input
states, for the case when the asymptotic state of the considered open system is a Gibbs state corresponding to two independent quantum harmonic oscillators in thermal equilibrium. The dynamics of the quantum entanglement strongly depends on the initial states and the parameters characterizing the environment (dissipation coefficient and temperature). For all values of the temperature of the thermal reservoir, an initial separable Gaussian state remains separable for all times. In the case of an entangled initial Gaussian state, entanglement suppression (entanglement sudden death) takes place for non-zero temperatures of the environment. Only for a zero temperature of the thermal bath the initial entangled state remains entangled for finite times, but in the limit of infinite time it evolves asymptotically to an equilibrium state which is always separable. The time when the entanglement is suppressed, decreases with increasing the temperature and dissipation.
We described also the time evolution of the logarithmic negativity, which characterizes the degree of entanglement of the quantum state, and it confirms the fact that the asymptotic equilibrium state is always separable.

\ack
I acknowledge the financial support received from the Romanian Ministry of Education and Research, through
the Projects IDEI 497/2009 and PN 09 37 01 02/2010.

\end{document}